\documentclass[11pt]{article}
\usepackage{moriond,epsfig}

\def\be{\begin{equation}}
\def\ee{\end{equation}}
\def\bea{\begin{eqnarray}}
\def\eea{\end{eqnarray}}

\def\met{\hbox{$\rlap{\kern0.25em/}E_T$}}

\begin{document}
\vspace*{4cm}
\title{TOP QUARK RESULTS FROM D\O}

\author{H.B. GREENLEE\\for the D\O\ Collaboration}

\address{Fermilab, P.O. Box 500, Batavia, IL, 60510, USA}

\maketitle\abstracts{
In this talk I will present recent preliminary results from the D\O\
experiment from Tevatron Run II ($p\bar p$ collisions at $\sqrt{s} =
1.96$ TeV).  The results presented in this talk include top quark pair
production cross section, top quark mass, and upper limits on single
top quark production.  }

The top quark was discovered at the Tevatron in 1995 by the CDF and
D\O\ experiments.\cite{disc} The top quark is by far the heaviest of
the six known quarks ($m_t = 178.0\pm4.3$ GeV).\cite{topmass} It is
also the least well studied.  The current run of the
Tevatron (Run II) will greatly increase the statistics availble for
studying the top quark and should result in many new and improved
experimental measurements.  The top quark was discovered using data
samples with in integrated luminosity from each experiment of about 50
pb$^{-1}$ and consisting of a few to a few dozen candidate events,
depending on channel.  The results presented in this talk are based on
integrated luminosities of about 200 pb$^{-1}$ out of about 500
pb$^{-1}$ recorded.  The balance of Tevatron Run II should increase
the integrated luminosity by a further order of magnitude to $>5$
fb$^{-1}$.

The Tevatron is a proton-antiproton collider with a center-of-mass
energy of $\sqrt{s} = 1.96$ TeV for Run II (as compared to 1.8 TeV for
Run I, during which run the top quark was discovered).  At hadron
colliders such as the Tevatron, top quarks can be produced in pairs by
the strong interaction, or singly by the weak interaction.  Perhaps
surprisingly, the expected pair production cross section
($\sigma_{t\bar t} \sim \hbox{6.8 pb}$)\cite{pairxsec} is only about
a factor of two larger than the expected single production cross
section ($\sigma_t
\sim 2.9$ pb).\cite{singlexsec}  
However, top quarks produced as pairs are much easier to observe than
singly produced top quarks due to lower background.  
The goal of top
quark research at the Tevatron is to test the Standard Model
predictions regarding the top quark by observing the top quark in as
many channels as possible, and to measure the fundamental properties
of the top quark, such as its mass and couplings.

\section{Top Quark Pair Production Cross Section}

The Standard Model predicts that top quarks should decay nearly 100\%
of the time into a $W$ boson and a $b$ quark.  Top quark pair events
are classified according to the decays of the two $W$ bosons as
dilepton events ($ee$, $e\mu$, $\mu\mu$, BR=5\%), lepton+jets events
($e$+jets, $\mu$+jets, BR=30\%), all jets events (BR=44\%), and events
containing taus ($e\tau$, $\mu\tau$, $\tau\tau$, $\tau$+jets,
BR=21\%).  Note that regardless of how the $W$ bosons decay, two $b$
quark jets are included in the final state, which is an important part
of the signature of top quark production.

\subsection{Dilepton Channel}

The dilepton channel signature is two high-$p_T$ isolated leptons, two
jets, and missing $E_T$ ($\met$).  The main backgrounds are $WW$,
$Z\to\tau\tau$, $Z\to ee$ and $Z\to\mu\mu$ (with missing $E_T$ due to
measurement errors), and backgrounds involving misidentified charged
leptons (fake leptons) in $W$+jets and QCD multijet events.  The event
selection cuts are as follows.
\begin{itemize}
\item
Lepton $p_T > 15$ GeV (20 GeV for $ee$).
\item
Jet $E_T > 20$ GeV.
\item
$\met > 25$ GeV ($e\mu$) or 35 GeV ($ee$ and $\mu\mu$).
\item
Rejection of $ee$ and $\mu\mu$ events with invariant mass close to $M_Z$.
\item
$H_T^{\hbox{leading lepton}} > 120$ GeV ($\mu\mu$) or 140 Gev ($e\mu$).
\end{itemize}
$H_T^{\hbox{leading lepton}}$ is defined as the sum of the $E_T$'s of the
jets plus the $p_T$ of the higher $p_T$ lepton.

\begin{table}
\caption{\label{tbl:dilep}Expected and observed events and backgrounds in
the dilepton channels.}
\begin{tabular}{|c|ccc|c|}
\hline
& $ee$ & $\mu\mu$ & $e\mu$ & Total \\
\hline
Integrated luminosity (pb$^{-1}$) & 156 & 140 & 143 & \\
\hline
$Z$ & $0.15\pm0.10$ & $2.04\pm0.49$ & $0.47\pm0.17$ & $2.66\pm0.53$ \\
$WW$ & $0.14\pm0.08$ & $0.10\pm0.04$ & $0.29\pm0.06$ & $0.53\pm0.11$ \\
Fake leptons & $0.91\pm0.30$ & $0.46\pm0.20$ & $0.19\pm0.06$ & $1.56\pm0.36$ \\
\hline
Total background & $1.20\pm0.33$ & $2.61\pm0.53$ & $0.95\pm0.19$ & $4.76\pm0.65$ \\
Expected signal & $1.39\pm0.19$ & $0.83\pm0.15$ & $3.77\pm0.44$ & $5.99\pm0.50$ \\
Total expected & $2.59\pm0.38$ & $3.44\pm0.55$ & $4.73\pm0.49$ & $10.76\pm0.83$ \\
\hline
Observed & 5 & 4 & 8 & 17 \\
\hline
\end{tabular}
\end{table}

Results from the dilepton channels are summarized in
Table~\ref{tbl:dilep}.  The D\O\ top quark
pair production cross section preliminary result in the dilepton
channel is as follows.\cite{dilep}
\begin{equation}
\sigma_{t\bar t} = 14.3_{-4.3}^{+5.1}\hbox{ (stat.) }
_{-1.9}^{+2.6}\hbox{ (syst.) } \pm 0.9\hbox{ (lum.) pb.}
\end{equation}


\subsection{Lepton+Jets Channel}

The nominal lepton+jets channel signature is one high-$p_T$ isolated
lepton, four jets (including two $b$-quark jets), and missing $E_T$.
The major backgrounds are $W$+jets and QCD multijet events with fake
leptons.  The strategy for identifying top quark pair events in this
channel is to begin with preselection cuts that require the lepton,
jets, and missing $E_T$, followed by additional analysis based on
$b$-tagging or topological event shape variables.  The integrated
luminosities of the data samples used in these channels varied from
141 pb$^{-1}$ to 169 pb$^{-1}$.  Preselection cuts are as follows:
\begin{itemize}
\item
One isolated lepton with $p_T > 20$ GeV and pseudorapidity $|\eta| <
1.1$ ($e$) or $|\eta| < 2.0$ ($\mu$).
\item
$\met > 20$ GeV ($e$) or $\met > 17$ GeV ($\mu$).
\item
Four or more jets (three in $b$-tagged analysis) with $E_T > 15$ GeV
and $|\eta| < 2.5$.
\end{itemize}

\begin{figure}
\centerline{
\includegraphics[width=6in]{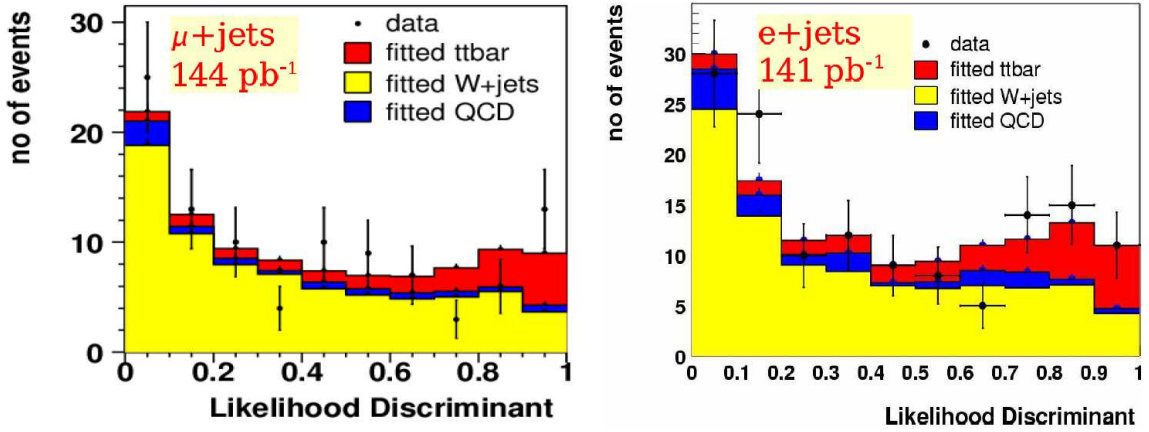}
}
\caption{\label{fig:topol}Topological lepton+jets cross section 
fit of data to sum of signal and background templates.}
\end{figure}

For the topological analysis, a likelihood discriminant is constructed
from four topological event shape variables (two of the variables are
sphericity and aplanarity).  Next, the cross section is calculated
using a maximum likelihood fit of the observed distribution of the
likelihood discriminant to the sum of signal and background templates
derived from either Monte Carlo (top quark signal, $W$+jets
background), or data (QCD multijet background).  For this fit, the
expected number of QCD multijet background is fixed, but the $W$+jets
contribution is allowed to float.  The results of the likelihood fit
are shown in Fig.~\ref{fig:topol}.  The preliminary cross section
result for the topological lepton+jets channel is
\begin{equation}
\sigma_{t\bar t} = 7.2_{-2.4}^{+2.6}\hbox{ (stat.) }
_{-1.7}^{+1.6}\hbox{ (syst.) } \pm 0.5\hbox{ (lum.) pb.}
\end{equation}

The second approach to extracting the top quark signal from
preselected $W$+jets events is based on $b$-tagging.  In this
approach, jets induced by $b$ quarks are identified by the presence of
displaced tracks arising from the decay of long-lived $b$-hadrons
(lifetime $b$-tagging) using the D\O\ silicon microstrip tracker
(SMT).  Two lifetime $b$-tagging algorithms are used, based on either
the impact parameters of tracks, or the presence of reconstructed
displaced secondary vertices.

\begin{figure}
\centerline{
\includegraphics[width=3in]{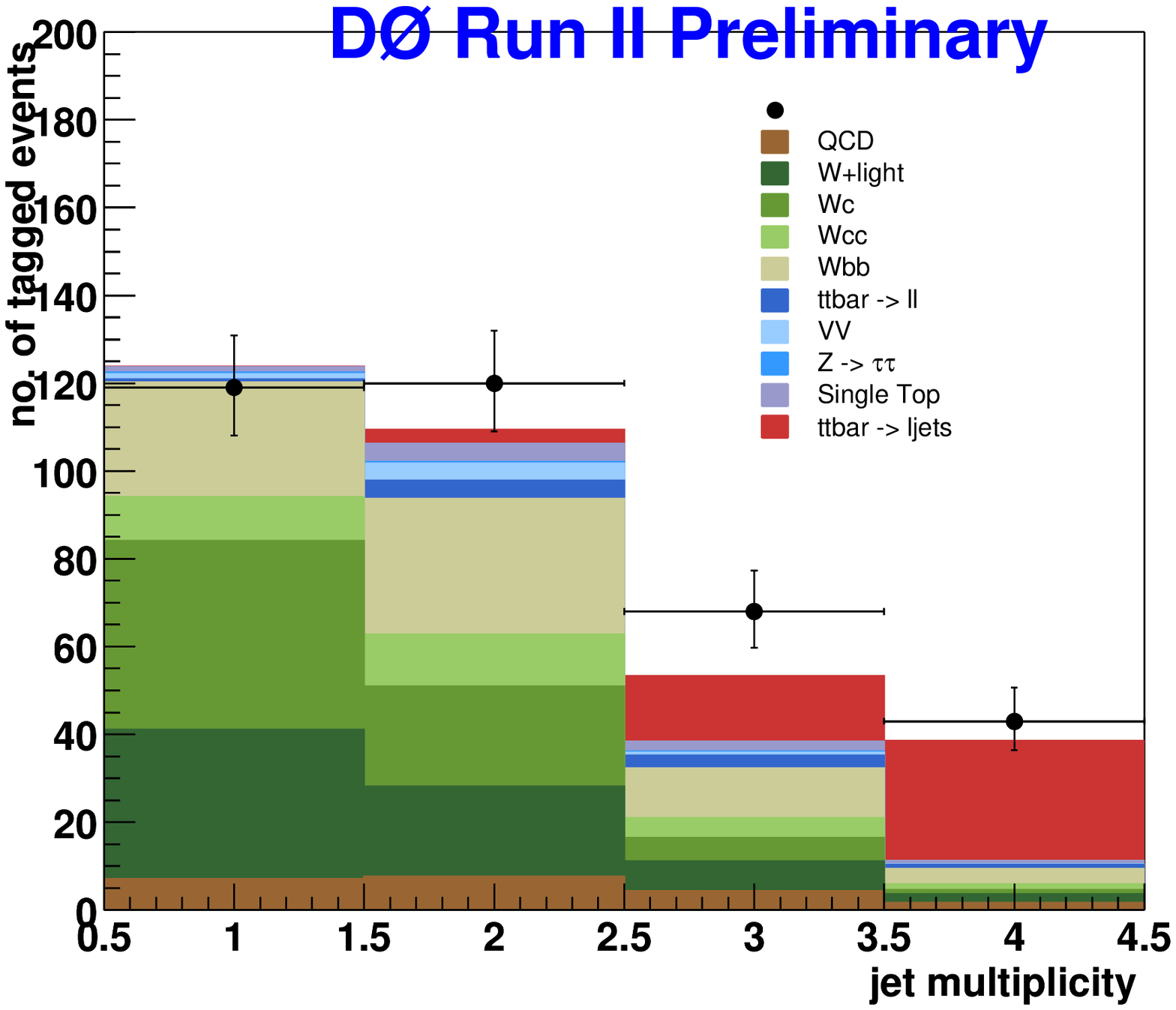}
\includegraphics[width=3in]{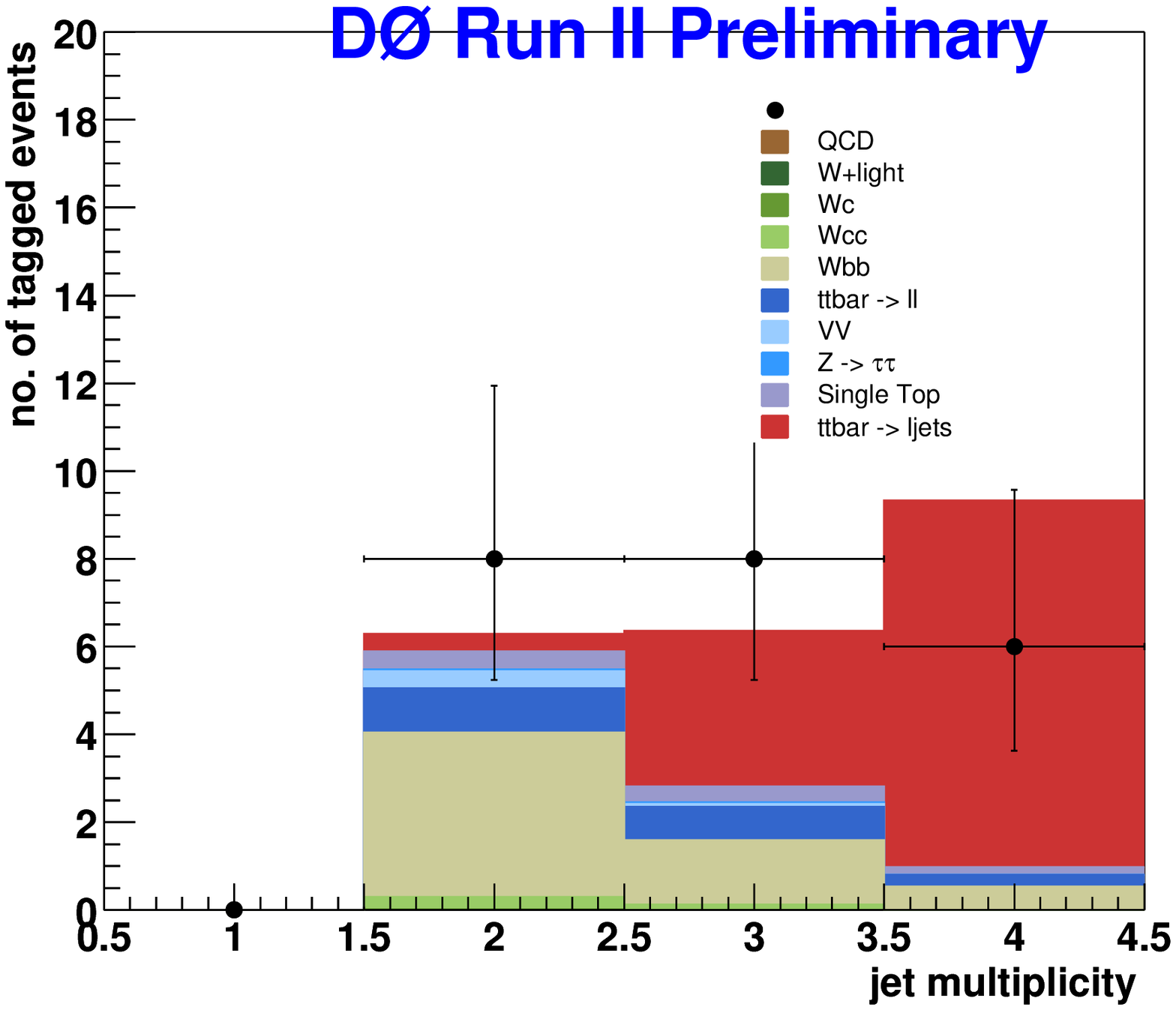}
}
\caption{\label{fig:btag}Jet multiplicity distribution for single and double
$b$-tagged $W$+jets events.}
\end{figure}

A likelihood fit is used to extract the top quark cross section.  For
the purpose of this fit, data are divided into separate bins for
electron vs.\ muon, three vs.\ four or more jets, and one vs.\ two or
more $b$-tagged jets (for a total of eight bins).  The expected
background normalization is fixed using known $b$-tagging efficiencies
and fake rates, and using a jet flavor composition estimated by Monte
Carlo.  The result of the likelihood fit is shown in
Fig.~\ref{fig:btag}.  The preliminary cross section results for the
$b$-tagged lepton+jets channel for the two $b$-tagging algorithms 
are\cite{btag}

\begin{equation}
\hbox{Impact parameter: }\sigma_{t\bar t} = 7.2_{-1.2}^{+1.3}\hbox{ (stat.) }
_{-1.4}^{+1.9}\hbox{ (syst.) } \pm 0.5\hbox{ (lum.) pb}.
\end{equation}
\begin{equation}
\hbox{Secondary vertex: }\sigma_{t\bar t} = 8.2_{-1.3}^{+1.3}\hbox{ (stat.) }
_{-1.6}^{+1.9}\hbox{ (syst.) } \pm 0.5\hbox{ (lum.) pb}.
\end{equation}

\subsection{All Jets Channel}

The nominal signature for the all jets channel is six jets, two of
which are $b$-quark jets.  The major background is QCD multijet
production.  The analysis of the all jets channel combines $b$-tagging
and event shape.  The integrated luminosity used in this analysis is
162 pb$^{-1}$.  The preselection cuts are as follows.
\begin{itemize}
\item
Six or more jets with $E_T > 15$ GeV and $|\eta| < 2.5$.
\item
One secondary vertex $b$-tag.
\end{itemize}

\begin{figure}
\centerline{
\includegraphics[width=3in]{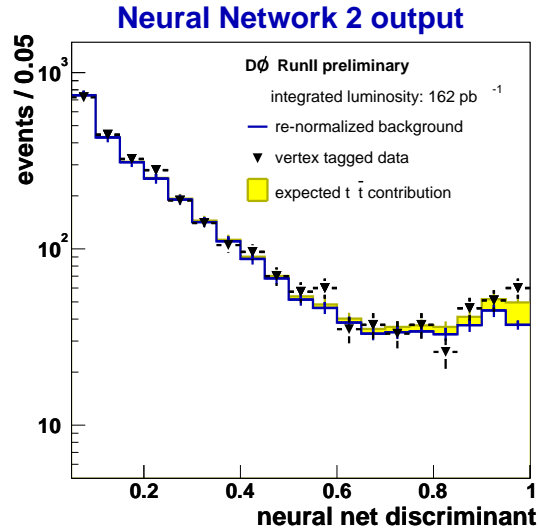}
}
\caption{\label{fig:alljets}Neural network output distribution for 
all jets channel.}
\end{figure}

The event shape analysis makes use of three neural networks constructed using
13 event shape variables.  The distribution of the final neural
network output is shown in Fig.~\ref{fig:alljets}.  The cross section
is calculated by making a cut on the neural network output ($NN >
0.75$), giving 220 events with an expected background of $186\pm5$.
The preliminary cross section result is\cite{alljets}
\begin{equation}
\sigma_{t\bar t} = 7.7_{-3.3}^{+3.4}\hbox{ (stat.) }
_{-3.8}^{+4.7}\hbox{ (syst.) } \pm 0.5\hbox{ (lum.) pb.}
\end{equation}

\begin{figure}
\centerline{
\includegraphics[width=3in]{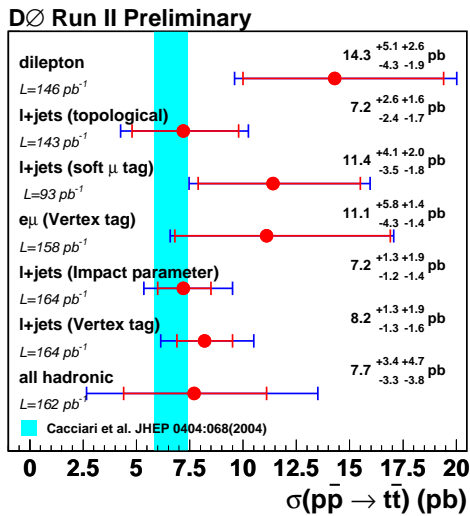}
}
\caption{\label{fig:cssum}D\O\ top quark 
pair production cross section results.}
\end{figure}

\subsection{Pair Production Summary}

Figure~\ref{fig:cssum} shows a summary of all D\O\ preliminary pair
production cross section results (those presented in this talk, and
others).  All results are consistent with each other and with theory.

\section{Top Quark Mass}

This talk contains top quark mass results from top quark pair events in the
dilepton and lepton+jets channels.  All results make use of some form of
template-based likelihood fit.  That is, a mass estimator is extracted for
each event, and the distribution of mass estimators is fit to the sum
of signal and background templates.  The top quark mass hypothesis is varied
to produce a likelihood curve.

\subsection{Top Quark Mass in the Dilepton Channel}

Dilepton top quark events do not have enough information to be fully
reconstructed.  Most dilepton events are therefore consistent with a
continuum of top quark masses.  It turns out that if one assumes a top
quark mass, then there is just enough information to fully reconstruct
dilepton events.  Not all top quark masses are equally likely,
however.  For any top quark mass hypothesis, events are assigned a
weight based on the Dalitz-Goldstein-Kondo (DGK) likelihood, which is
proportional to the differential cross section of the fully
reconstructed event.  In case there is more than one solution to the
reconstruction of the event (there can be up to four), the DGK
likelihood is the sum of the possible solutions.  Resolution smearing
is also taken into account in constructing the DGK likehood curve.
The DGK likelihood curve is used to extract a single top quark mass
estimator, which is the top quark mass which maximizes the DGK
likelihood.  This is not the only possible mass estimator, nor is it
the optimal one, but it is the simplest.

\begin{figure}
\centerline{\includegraphics[width=3in]{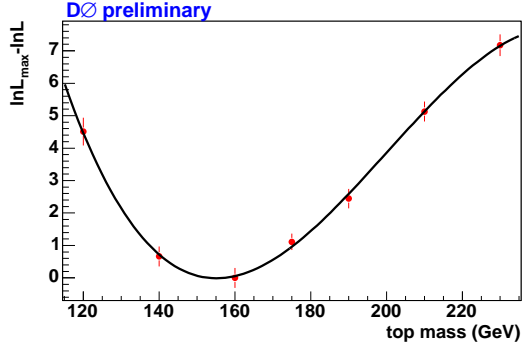}}
\caption{\label{fig:dilepmass}Dilepton top quark mass likelihood.}
\end{figure}

The event sample for the dilepton mass analysis consists of 13 events
(8 $e\mu$, 5 $ee$, 0$\mu\mu$), with an expected background of 3.3
events (1.0 $e\mu$, 0.9 $ee$, 1.4 $\mu\mu$).  
The top mass likelihood curve extracted from the 
signal and background template fit is shown in Fig.~\ref{fig:dilepmass}.
The preliminary result for the top quark mass in the dilepton channel is
as follows.\cite{dilepmass}
\begin{equation}
m_t = 155_{-13}^{+14}\hbox{ (stat.) } \pm 7\hbox{ (syst.) } \hbox{ GeV}/c^2.
\end{equation}


\subsection{Top Quark Mass in the Lepton+Jets Channel}

Candidate events for the lepton+jets mass analysis are preselected
as $W$+4 jet events, with final selection based either on topological
event shape or $b$-tagging (similar to the cross section analysis).

\begin{figure}
\centerline{
\includegraphics[width=3in]{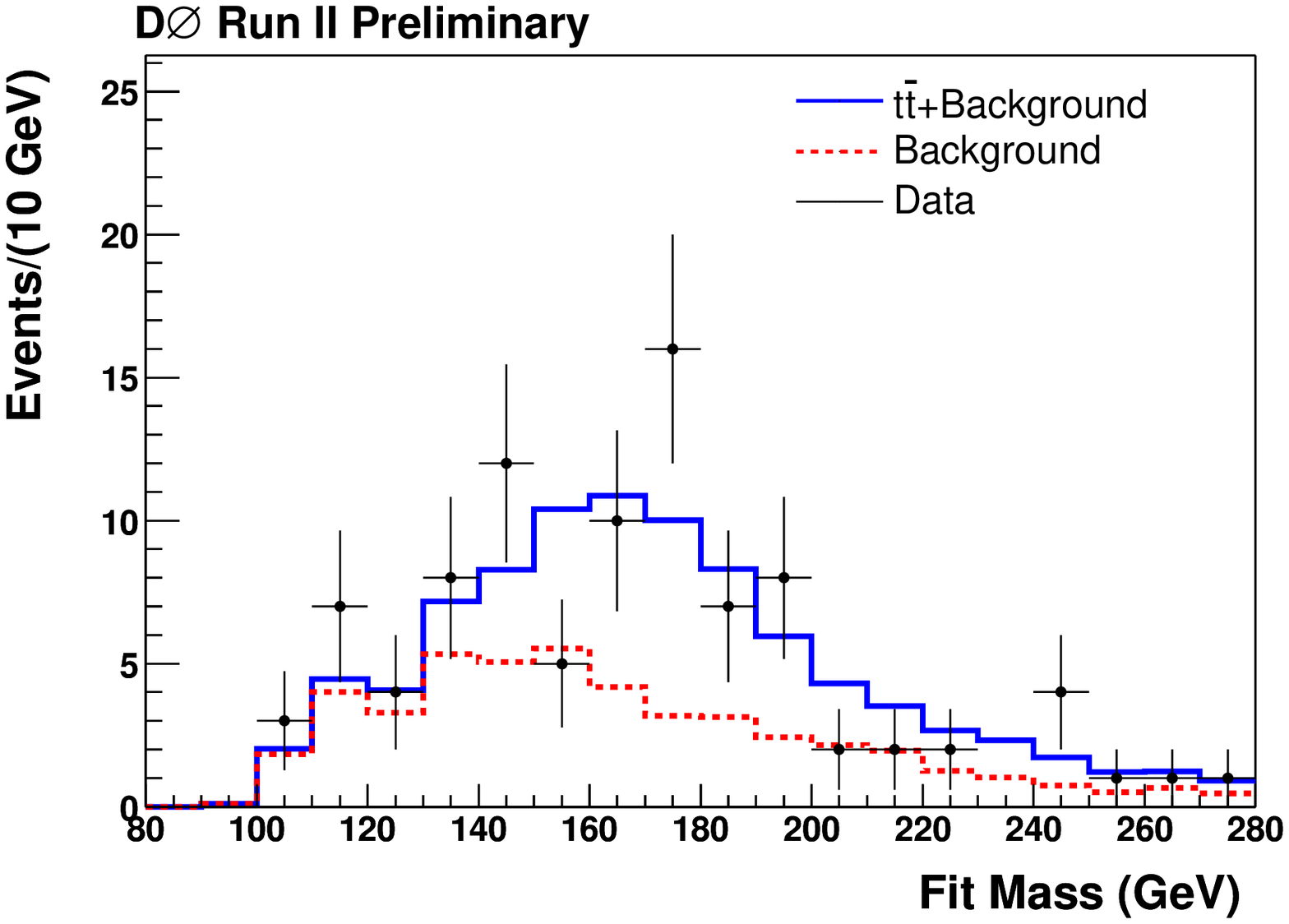}
\includegraphics[width=3in]{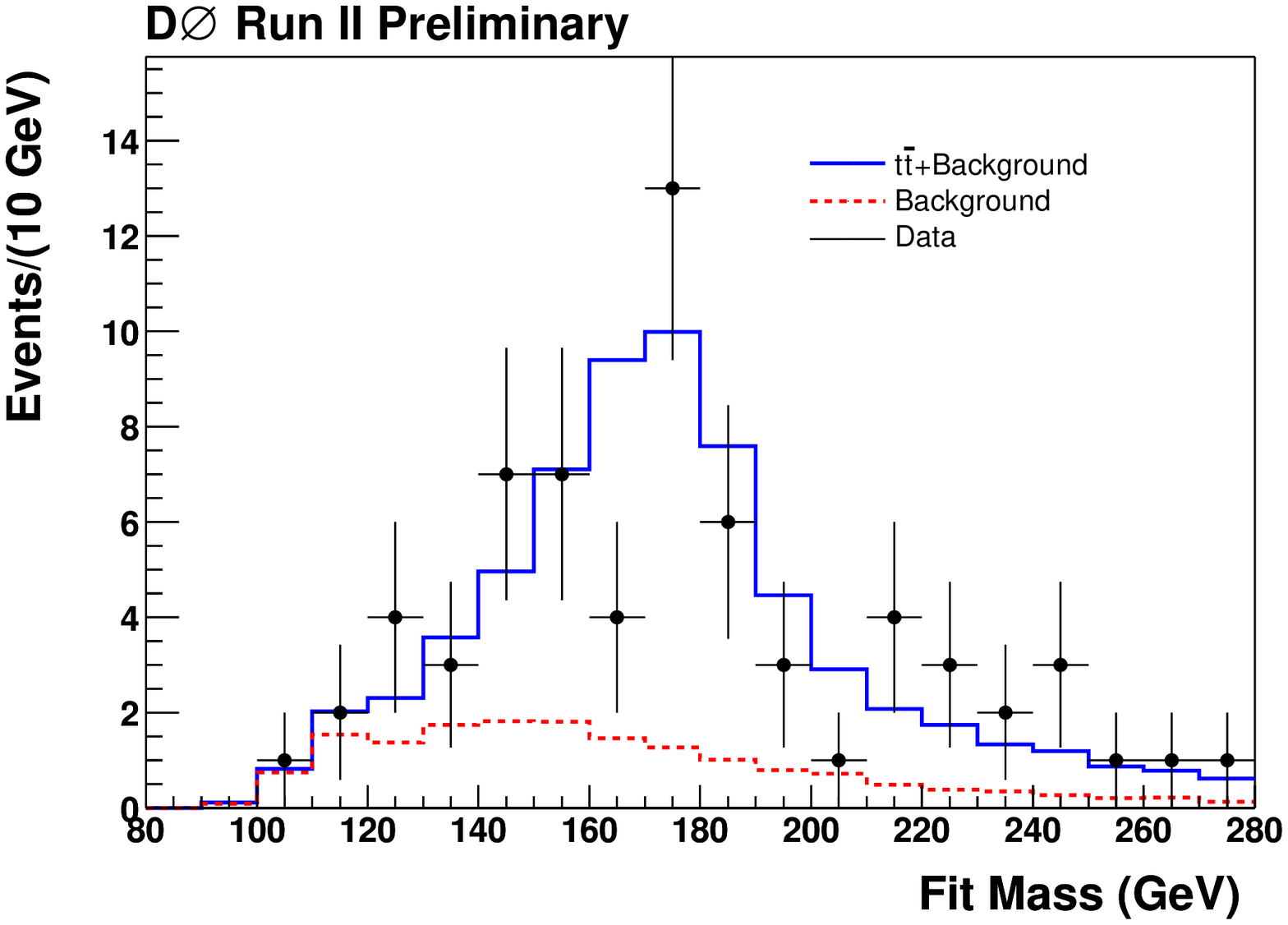}
}
\caption{\label{fig:ljetsmass}Lepton+jets top quark mass likelihood fit
for untagged events (left) and $b$-tagged events (right).}
\end{figure}

The top quark mass is estimated for each event by fully reconstructing
the the event according to the top quark pair hypothesis using a
kinematic fit.  The kinematic fit has 12 possible solutions (six with
single $b$-tag, two with double $b$-tag).  The top quark mass estimate
is taken from the best fit.  The results of the likelihood fit are shown
in Fig.~\ref{fig:ljetsmass}.  The preliminary top quark mass in the
lepton+jets channel for untagged and $b$-tagged events are\cite{ljetsmass}
\begin{equation}
\hbox{Untagged: }m_t = 169.9\pm 5.8\hbox{ (stat.) }
_{-7.1}^{+7.8}\hbox{ (syst.) } \hbox{ GeV}/c^2,
\end{equation}
\begin{equation}
\hbox{$b$-tagged: }m_t = 170.6\pm 4.2\hbox{ (stat.) } 
\pm 6.0\hbox{ (syst.) } \hbox{ GeV}/c^2,
\end{equation}

\section{Single Top Production}

Production of top quarks singly by the weak interaction provides the
opportunity to make an absolute measurement of the weak couplings of
the top quark (e.g. $|V_{tb}|$), which is not possible with
pair-produced top quark events.  Observing single top quark production
is an experimental challenge, which may be quite close to being
achieved.

The leptonic signature for single top production is similar to the
lepton+jets channel for top pairs, except that there are fewer jets.
The final state contains two $b$-quarks, one of which is produced in
association with the top quark (which may or may not too far forward
to be observable), and the second of which comes from top quark
decay.  Major backgrounds are $W$+jets, top quark pairs, and QCD
multijet events with fake leptons.  Event selection cuts for the
single top analysis are as follows.
\begin{itemize}
\item
One isolated lepton with $p_T>15$ GeV$/c$ and $|\eta|<1.1$ ($e$) or
$|\eta| < 2.0$ ($\mu$).
\item
$\met > 15$ GeV.
\item
Leading jet $E_T > 25$ GeV and $|\eta| < 2.5$.
\item
One to three additional jets with $E_T > 15$ GeV and $|\eta| < 3.4$.
\item
One or two secondary vertex $b$-tags.
\end{itemize}

\begin{figure}
\centerline{
\includegraphics[width=3in]{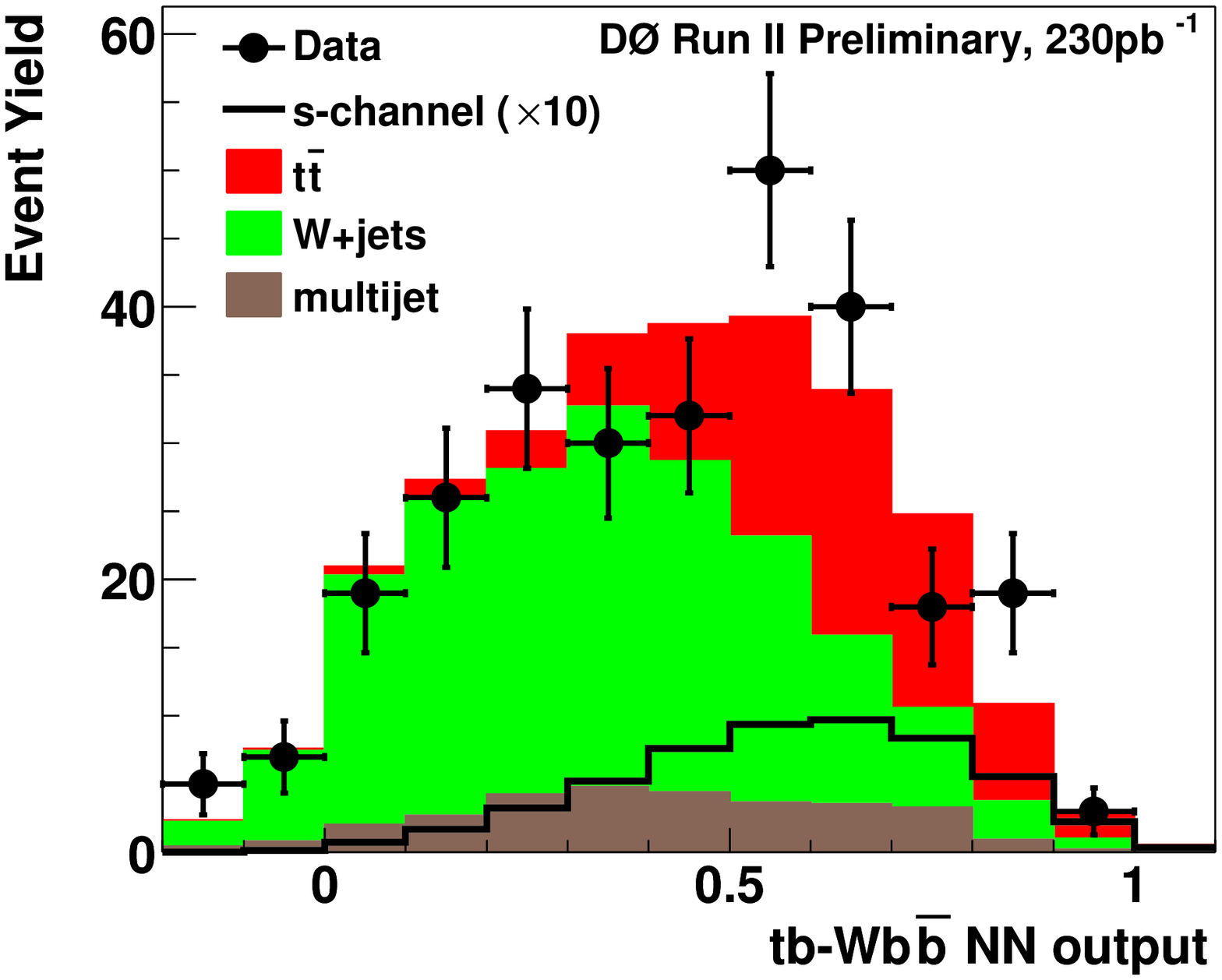}
\includegraphics[width=3in]{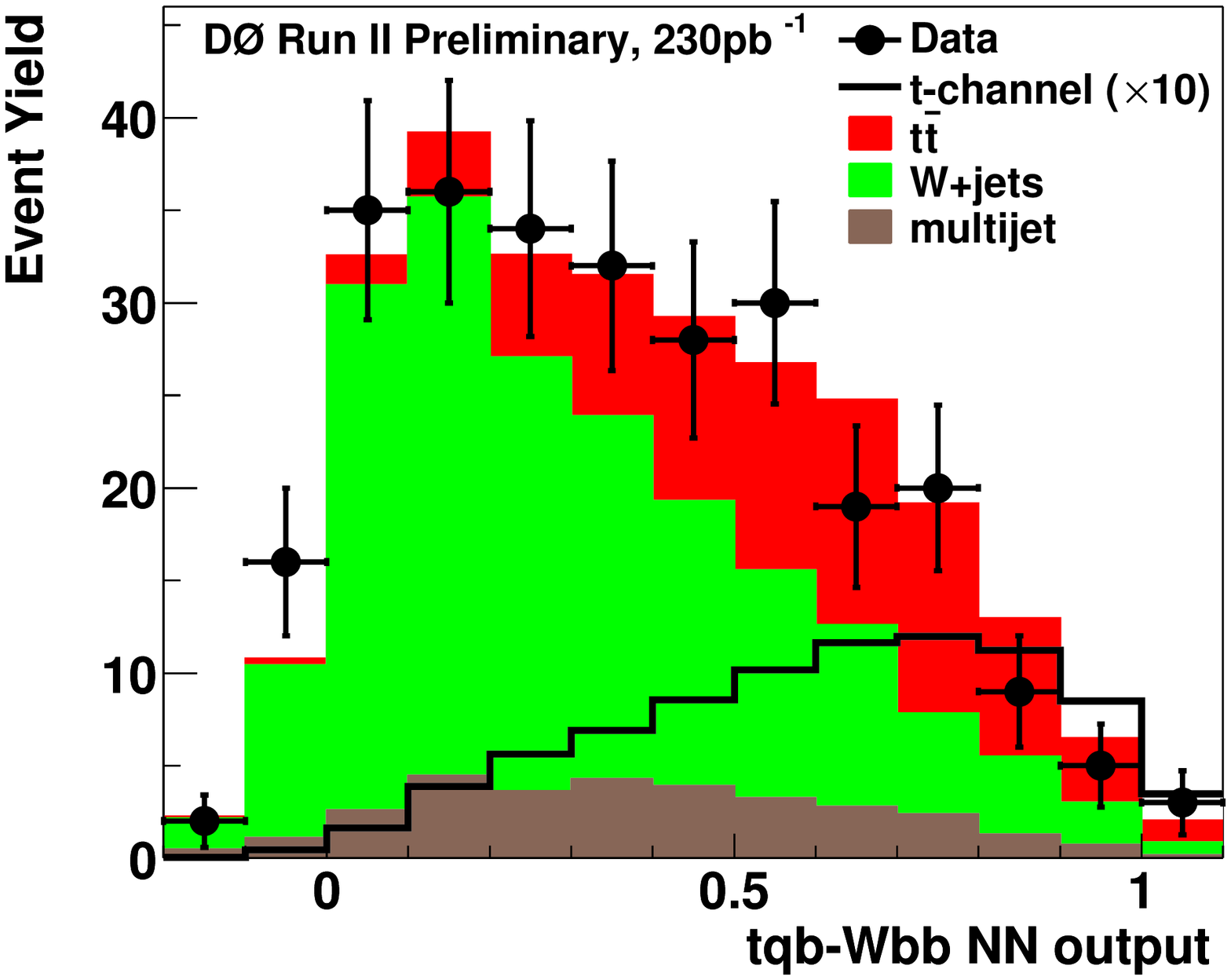}
}
\centerline{
\includegraphics[width=3in]{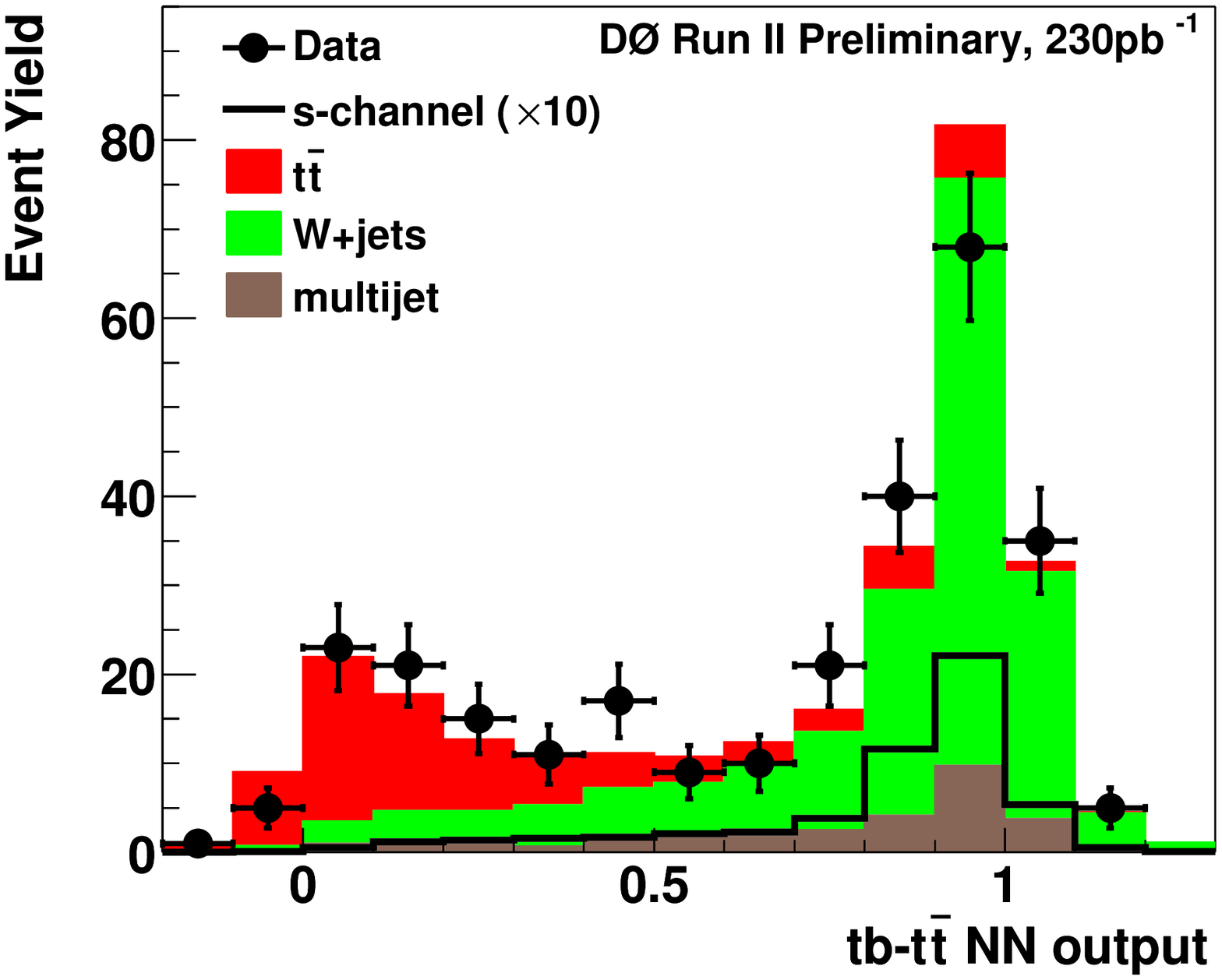}
\includegraphics[width=3in]{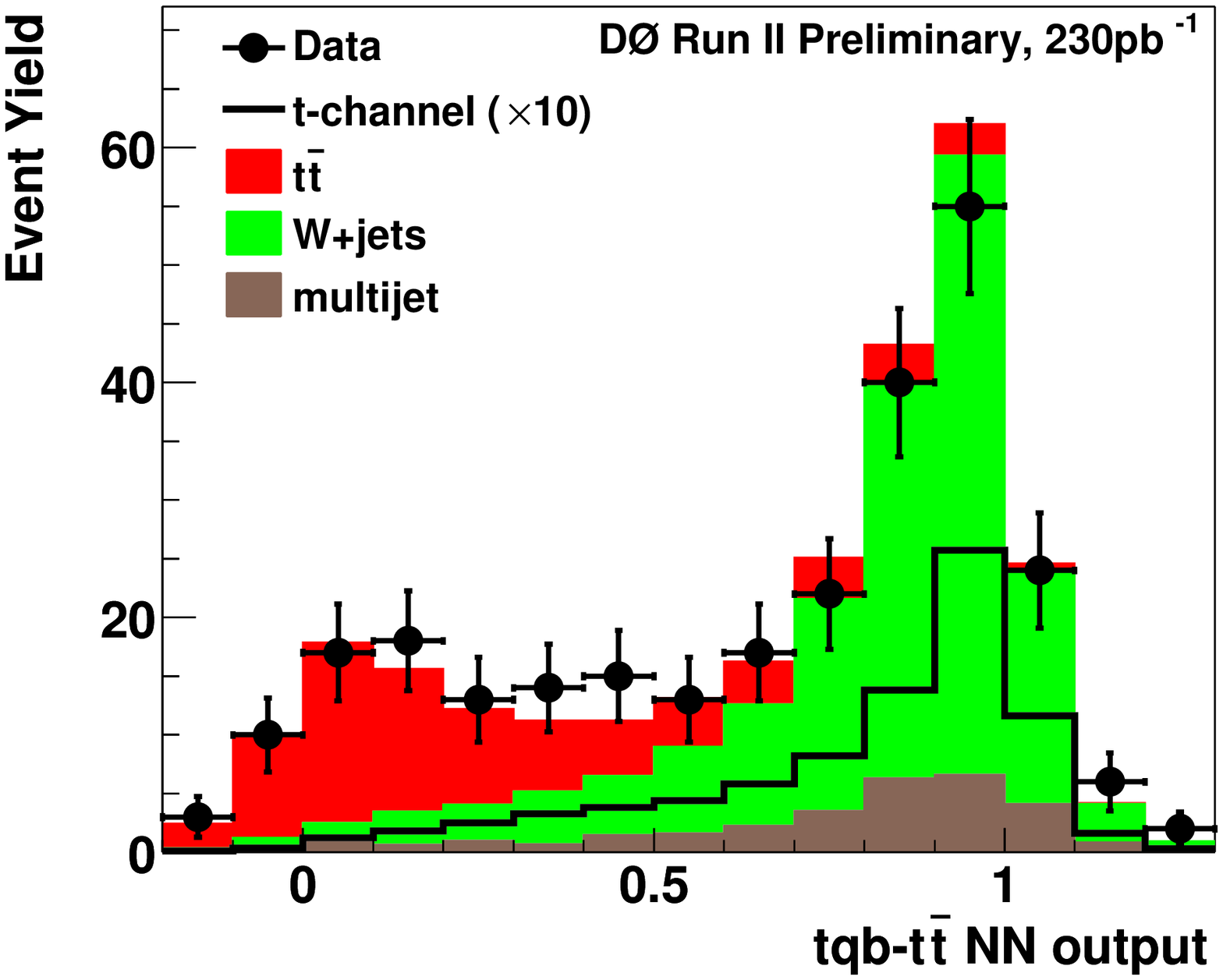}
}
\caption{\label{fig:stnn}Distributions of single top neural network
discriminants.}
\end{figure}

Following event selection, a neural network event shape analysis is
performed using 11 event shape variables.  These variables include
energy-related variables, including the reconstructed top quark mass,
as well as non-energy-related variables.  Four neural network
discriminants are derived, one of which is optimized for separating
each of the two signal production mechanisms ($s$-channel and
$t$-channel) from each of the two main backgrounds ($W$+jets and
$t\bar t$).  Distributions of the four neural network discriminants
are shown in Fig.~\ref{fig:stnn}.

\begin{figure}
\centerline{
\includegraphics[width=3in]{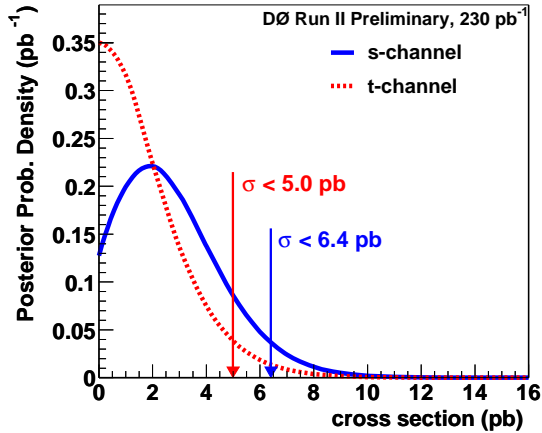}
}
\caption{\label{fig:stlimit}Likelihood curve sof $s$-channel and $t$-channel
single top production.}
\end{figure}

Upper limits are obtained for each production mechanism separately
using a two-dimsneional template likelihood fit, whose two
dimensions are the neural network discriminants for the two
backgrounds.  The likelihood curve for the two production mechanisms
are shown in Fig.~\ref{fig:stlimit}.  The preliminary results for the
upper limits on single top production are as follows.\cite{singletop}

\begin{equation}
\hbox{$s$-channel: }\sigma_t < 6.4\hbox{ pb (95\% CL)}
\end{equation}
\begin{equation}
\hbox{$t$-channel: }\sigma_t < 5.0\hbox{ pb (95\% CL)}
\end{equation}
The expected limits are 4.5 pb and 5.8 pb for $s$-channel and $t$-channel
respectively.

\section{Summary}

To summarize, in this talk we present results for top quark pair production,
top quark mass, and upper limits on single top quark production.  All results
are consistent with previous results and with the Standard Model.  All results
are preliminary, based on integrated luminosities of about 200 pb$^{-1}$.
Tevatron Run II is expected to generate data with an integrated luminosity
in the range 4--8 pb$^{-1}$.

\section*{Acknowledgments}
%
We thank the staffs at Fermilab and collaborating institutions, 
and acknowledge support from the 
DOE and NSF (USA);
CEA and CNRS/IN2P3 (France);
FASI, Rosatom and RFBR (Russia);
CAPES, CNPq, FAPERJ, FAPESP and FUNDUNESP (Brazil);
DAE and DST (India);
Colciencias (Colombia);
CONACyT (Mexico);
KRF (Korea);
CONICET and UBACyT (Argentina);
FOM (The Netherlands);
PPARC (United Kingdom);
MSMT (Czech Republic);
CRC Program, CFI, NSERC and WestGrid Project (Canada);
BMBF and DFG (Germany);
SFI (Ireland);
Research Corporation,
Alexander von Humboldt Foundation,
and the Marie Curie Program.

\end{document}